# Near-ultraviolet photon-counting dual-comb spectroscopy

Bingxin Xu [1], Zaijun Chen [1,#], Theodor W. Hänsch [1,2], Nathalie Picqué [1,*]

[1] Max-Planck Institute of Quantum Optics, Hans-Kopfermann-Straße 1, 85748, Garching, Germany
[2] Ludwig-Maximilian University of Munich, Faculty of Physics, Schellingstr. 4/III, 80799, München, Germany
# Present address: Ming Hsieh Department of Electrical and Computer Engineering, University of Southern California, Los Angeles, California 90089, USA
* Corresponding author. email: nathalie.picque@mpq.mpg.de, http://www.frequency-comb.eu

**Abstract**

Ultraviolet spectroscopy provides unique insights into the structure of matter with applications ranging from fundamental tests to photochemistry in the earth's atmosphere and astronomical observations from space telescopes. At longer wavelengths, dual-comb spectroscopy with two interfering laser frequency combs has evolved into a powerful technique that can offer simultaneously a broad spectral range and very high resolution. Here we demonstrate a photon-counting approach that can extend the unique advantages of this method into ultraviolet regions where nonlinear frequency-conversion tends to be very inefficient. Our spectrometer, based on two frequency combs of slightly different repetition frequencies, provides broad span, high resolution, frequency calibration within the accuracy of an atomic clock, and overall consistency of the spectra. We demonstrate a signal-to-noise ratio at the quantum limit and optimal use of the measurement time, provided by the multiplex recording of all spectral data on a single photo-counter. Our initial experiments are performed in the near-ultraviolet and in the visible spectral ranges with alkali-atom vapor, with a power per comb line as low as a femtowatt. This crucial step towards precision broadband spectroscopy at short wavelengths clears the path to extreme-ultraviolet dual-comb spectroscopy and, more generally, generates a new realm of applications for diagnostics at photon level, as encountered e.g., when driving single atoms or molecules.

**Introduction**

We take a significant step towards precise spectroscopy over broad spectral bandwidths in the extreme ultraviolet spectral region by demonstrating a novel approach to dual-comb spectroscopy that is suited to extremely low light levels, as encountered in frequency-comb photonics at short wavelengths. By utilizing photon-counting technology, we have achieved precise, high-resolution, quantum-noise-limited near-ultraviolet dual-comb spectroscopy that operates at photon fluxes up to $10^8$ lower than commonly employed levels. This breakthrough opens up novel opportunities for precise ultraviolet spectroscopy across a wide spectral bandwidth.

Ultraviolet spectroscopy plays a pivotal role in studying electronic transitions in atoms and rovibronic transitions in molecules, essential for tests of fundamental physics and of quantum-electrodynamics theory [1,2], determination of fundamental constants [3], precision measurements [4], optical clocks [5], high-resolution spectroscopy supporting atmospheric chemistry [6] and astrophysics [7,8], as well as strong-field physics [9].

The emerging dual-comb technique [10-22] offers spectroscopy a unique host of features, which would provide a unique tool for vacuum- and extreme-ultraviolet spectroscopy. Dual-comb spectroscopy leverages frequency combs, spectra of evenly spaced phase-coherent laser lines which have revolutionized time and frequency metrology [23,24]. Dual-comb spectra span over a broad spectral bandwidth and their frequency scale may be directly calibrated within the accuracy of an atomic clock, while, as the transitions are interrogated with narrow laser lines, the well-defined instrumental line-shape usually has a negligible contribution compared to that of the atomic or molecular profiles [10]. Dual-comb spectroscopy is also a complex technique that involves recording, over long durations, the time-domain interference between the two frequency combs with slightly different repetition frequencies. Dual-comb spectroscopy currently attracts tremendous interest in the infrared range.

Considerable progress has been achieved for generating frequency combs at short wavelengths and using them with techniques of narrow-band spectroscopy [1-5]. To date, high-resolution dual-comb spectroscopy with ultraviolet radiation has not been reported. Previous studies have demonstrated nonlinear spectroscopy in rubidium using two-photon excitation with combs at 384 THz (780 nm) [25,26]. Linear absorption spectroscopy is of significant interest in atmospheric science, and two proposals have discussed possible implementations based on high-power laser systems for reaching short wavelengths [27,28]. Currently, several research groups are pursuing different approaches for experimental implementation [29-32].

Here, we present the first demonstration of comb-line-resolved, high-resolution dual-comb absorption spectroscopy in the near ultraviolet region, achieving resolutions of 500 MHz and 200 MHz at the center frequency of 772 THz (388 nm). Notably, we introduce an innovative approach based on photon counting, which overcomes the challenges posed by low conversion efficiency. Moreover, it enables a quantum-noise-limited signal-to-noise ratio, particularly difficult to achieve in dual-comb spectroscopy. Our robust method for interferometry at low light levels lays a solid foundation for extending dual-comb spectroscopy to even shorter wavelengths.

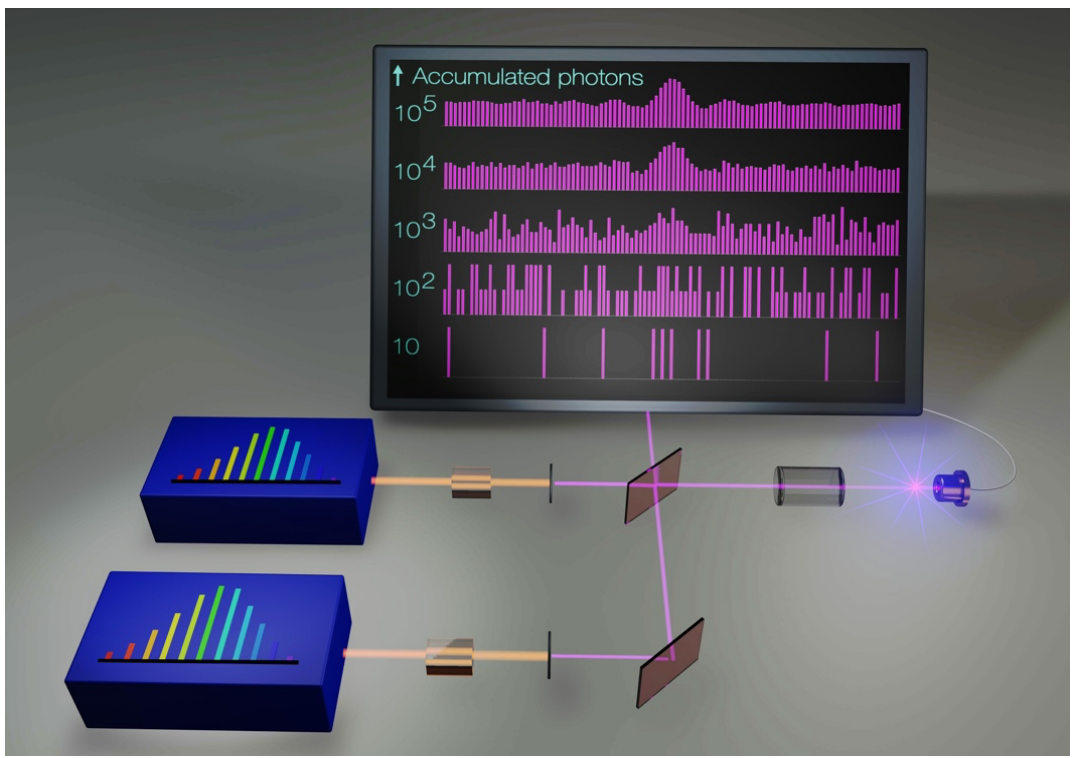

**Figure 1. Principle of ultra-violet dual-comb spectroscopy with photon counting.** The beams of two very-low-light-level comb generators of slightly different pulse repetition frequencies are superimposed with a beam splitter. One output is counted by a photon-counting detector. Less than one detector click occurs every twenty comb pulses. At power levels up to $10^8$ times weaker than usually employed in dual-comb spectroscopy, the statistics of the detected clicks carry the spectral information about the sample.

## Results

**Principle of photon-counting dual-comb spectroscopy**
We conduct dual-comb spectroscopy under extremely low light conditions, where, on average, fewer than one detector click occurs every twenty repetition periods of the comb. It is unlikely that two photons, one from each comb generator, would be present in the interferometer at the same time. We operate with optical powers that are more than a million times weaker than those commonly used in dual-comb spectroscopy. In [33], we had proposed the conceptual possibility of dual-comb spectroscopy in the photon-counting regime; however, its practical applicability has remained elusive, because of poor spectral resolution.

Two frequency comb generators emit, at very low light levels, trains of pulses at slightly different repetition frequencies, $f_{rep}$ and $f_{rep}+\delta f_{rep}$, respectively. The two beams of the two combs are combined on a beam splitter and their time-domain interference is measured on a fast photon counter. We record the detector clicks as a function of time during a specific duration, that we call a scan. A scan is initiated by a trigger signal generated in the instrument. Subsequent trigger signals repeat the same scans and add the detector clicks of a given delay into the already accumulated clicks of the same delay. Due to the low light level, it is crucial to accumulate multiple identical scans to reconstruct the interferograms using sufficient statistical data from photon counts. The sequence is repeated until the desired signal-to-noise ratio is achieved in the interferogram (Fig. 1). Direct accumulation is feasible if precise reproducibility of the interferometric scans is achieved. We accomplish this by using well-controlled frequency combs and experimental parameters which leads to reproducible interferometric waveforms. In particular, we set $\delta f_{ceo}$ = 0 [modulo $\delta f_{rep}$], where $\delta f_{ceo}$ is the difference in carrier offset frequencies of the two combs. The sampling rate of the time bins should be an integer multiple of the difference in repetition frequencies $\delta f_{rep}$. Here, we demonstrate accumulation times of over one hour at quantum-noise limited sensitivity.

Once the acquisition is completed with sufficient statistics, the interferogram can be processed similarly to dual-comb interferograms obtained at higher power. In the frequency domain, pairs of optical comb lines, one from each comb, produce radio-frequency beat notes on the detector, forming, in the radio-frequency domain, a frequency comb of line spacing $\delta f_{rep}$ and with a carrier-envelope offset frequency equal to zero. Optical frequencies are thus down-converted into radio frequencies, m $\delta f_{rep}$, where m is an integer.

**Near-ultraviolet experimental spectra with electro-optic combs**
We illustrate the potential of near-ultraviolet dual-comb spectroscopy by using nonlinear frequency conversion of near-infrared electro-optic frequency combs (Fig.1, Supplementary Fig. 1, Methods). The electro-optic system, with its poor conversion efficiency in the ultra-violet, is well suited to test our approach to photon counting. Two frequency combs of slightly different repetition frequencies are generated from a continuous-wave laser that is intensity- and phase-modulated by electro-optic modulators at a center frequency of 193 THz (1550 nm). An acousto-optic modulator offsets the center frequency of one comb, in order to measure the dual-comb spectrum without aliasing. The near-infrared combs are sequentially frequency-doubled twice, once in a periodically-poled lithium niobate crystal and once in a $BiB_3O_6$ crystal. The near-ultraviolet combs of about 100 lines have a center frequency tunable between 770 and 774 THz, and a repetition frequency which can be freely chosen between 100 kHz and 40 GHz. The two low-power ultraviolet frequency comb beams are combined on a beamsplitter. One output of the beamsplitter is detected by a photon counter. The detected photon rate of the combined beam at the photon counter is at most $5 \cdot 10^7$ photons s$^{-1}$. This corresponds to an average optical power per comb before the detector of about 50 pW, more than one-million-fold weaker than that commonly used in dual-comb spectroscopy. The counts are sampled by a multiscaler at a rate ranging between 12.5 GHz and 500 MHz. The trigger signal for the data acquisition by the multiscaler is produced by frequency division of the 10-MHz clock signal.

An experimental dual-comb interferogram sampled at a rate of 12.5 GHz (Supplementary Fig. 2) recurs at a period of 625 ns, owing to a difference in repetition frequency $\delta f_{rep}$ of 1.6 MHz. A Fourier transform reveals the dual-comb spectrum, as in traditional dual-comb spectroscopy. A dual-comb spectrum centred at 770.73 THz contains more than 100 comb lines with a flat-top intensity distribution and results from an accumulation time of 255.5 s (Fig.2a). The spectral resolution is equal to the line-spacing $f_{rep}$ of 500 MHz. The span is 50 GHz. The instrumental line shape is a cardinal sine, as expected in an unapodized spectrum (Fig. 2b). By tuning the centre frequency of the continuous-wave near-infrared laser, a sequence of dual-comb spectra corresponding to 15 acquisitions centred at different frequencies is acquired over a total span of about 3 THz (Fig. 2c).

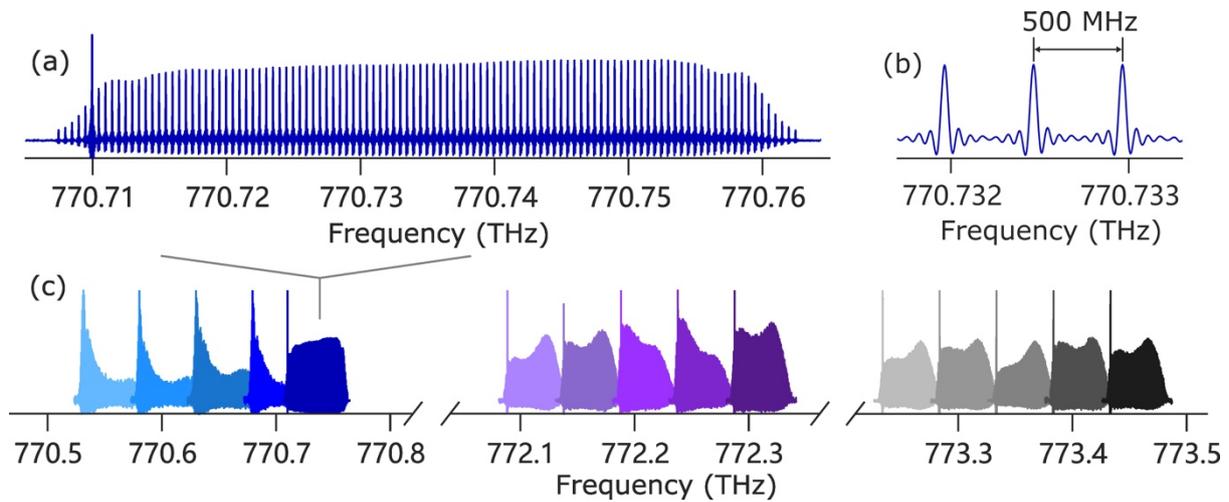

**Figure 2. Near-ultraviolet photon-level dual-comb experimental spectra with resolved comb lines.** (a) Spectrum recorded at a detected rate of $2.7 \; 10^7$ photons s$^{-1}$ with about 100 comb lines. The y-scale is linear. (b) Magnified portion of (a) showing 3 comb lines with cardinal-sine lineshapes. (c) Illustration of the frequency agility: a sequential acquisition of 15 spectra spanning overall a bandwidth broader than 3 THz.

Next, we illustrate the potential of our photon-counting dual-comb interferometer for absorption spectroscopy of the weak 6S-8P transitions in atomic caesium vapor. After the beam splitter, the combined comb beam passes through a heated caesium cell. The interferograms are detected at a rate of $3 \; 10^7$ counts s$^{-1}$, which corresponds to an average optical power per comb of $30 \; 10^{-12}$ W before the detector of a quantum efficiency of 25% (Table 1, Methods). With about 100 comb lines, the power per comb line is estimated to $3 \; 10^{-13}$ W, simply calculated as the optical power divided by the number of comb lines. Spectra with resolved comb lines exhibit the $6S_{1/2}$-$8P_{1/2}$ and $6S_{1/2}$-$8P_{3/2}$ transitions at 770.73 THz and 773.21 THz, respectively (Fig. 3a and Fig. 3c, respectively) [34]. The transmittance spectrum of the $6S_{1/2}$-$8P_{1/2}$ transitions of a Doppler full-width at half-maximum of 1 GHz results from an accumulation time of 267 s (Fig. 3b). The two resonances are due to the hyperfine splitting in the $6S_{1/2}$ ground state, while the hyperfine structure in the excited state is not resolved owing to the Doppler broadening. The signal-to-noise ratio, determined as the inverse standard deviation of the normalized absorption baseline of the amplitude spectrum, is 345. The stronger $6S_{1/2}$- 8P$_{3/2}$ transitions are measured at a lower temperature and shorter accumulation time (82 s) at a signal to noise ratio of 205 (Fig.3d).

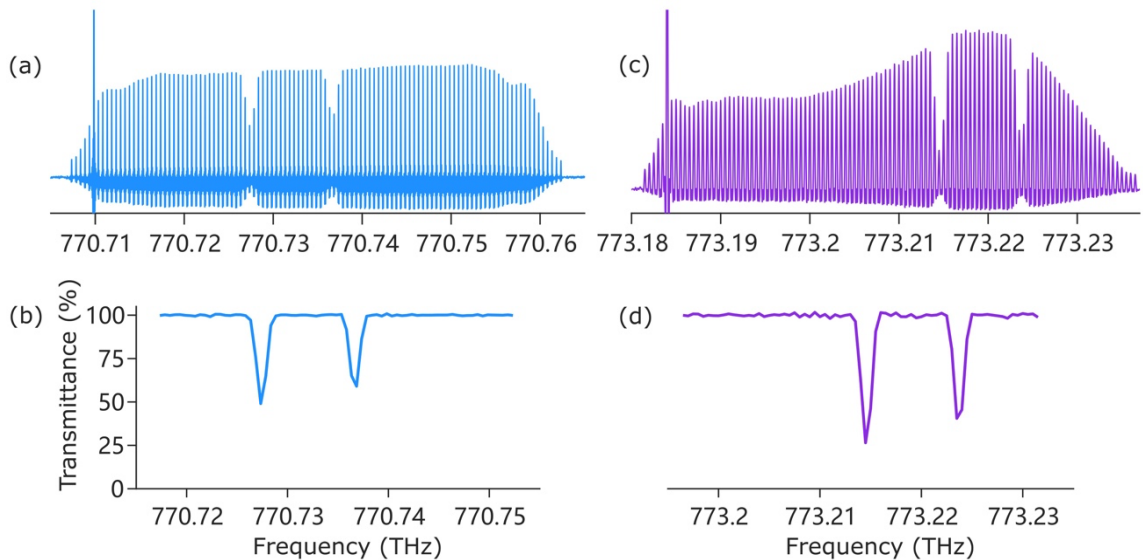

**Figure 3. Photon-counting near-ultraviolet spectra of weak transitions in caesium at a total average optical power of 60 pW and 500-MHz resolution.** The y-scale is linear. The spectra span over 50 GHz. (a)(b) 6S$_{1/2}$-8P$_{1/2}$ resonances with a signal-to-noise ratio of the transmittance baseline of 345. (c) 6S$_{1/2}$-8P$_{3/2}$ transitions with a signal-to-noise ratio of 205.

One may argue that each spectrum does not span over extended regions. This feature of dual-comb spectroscopy with electro-optic modulators has been discussed in the context of infrared spectroscopy [19]: the moderate spectral span is counter-weighted by the frequency agility of the continuous-wave laser and of the driving synthesizers: the center frequency and the repetition frequency of the combs can be adjusted by simply dialing a knob and the small number of comb lines provides a high signal-to-noise ratio in short measurement times, with selectable spectral resolution. This makes the setup appealing for some experimental configurations such as atomic spectroscopy where the spectra are not too dense and crowded. An illustration of the adjustable resolution compares the transmission spectrum of the 6S$_{1/2}$-8P$_{1/2}$ transition at a resolution of 500 MHz with that at a resolution of 200 MHz (Supplementary Fig. 3).

Reducing technical noise sources to reach the quantum-noise limit is not easy in dual-comb spectroscopy. Most experiments have been limited by detector noise or intensity noise of the laser sources. In the experimental conditions of Fig. 3a, we measure the experimental signal-to-noise ratio of the absorption baseline in the spectra as a function of count rate (Fig. 4a) and accumulation time (Fig. 4b). The experimental signal-to-noise ratio scales as the square-root of the detected photon rate (Fig. 4a), which demonstrates that the signal-to-noise ratio is limited by counting statistics (also called quantum-noise limit or shot-noise limit). The experimental signal-to-noise is also in good agreement with a simple model of the quantum-noise-limited signal-to-noise ratio (Methods). The experimental signal-to-noise also scales with the square-root of the accumulation time, showing that the interferometric coherence is maintained (Fig. 4b).

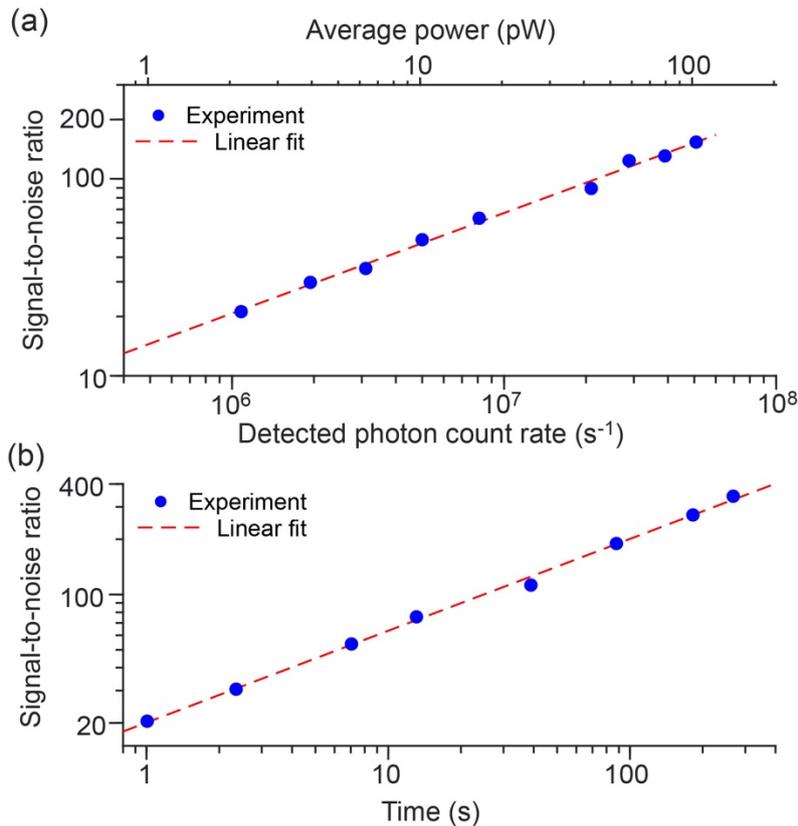

**Figure 4. Quantum-noise-limited signal-to-noise ratio in photon-counting near-ultraviolet dual-comb spectroscopy.** (a) The signal-to-noise ratio of the transmittance baseline scales as the square-root of the photon rate, showing that the quantum-noise limit is reached. The slope in the linear fit is 0.51 (+/- 0.01). (b) The signal-to-noise ratio also scales as the square-root of the accumulation time, showing that mutual coherence of the two combs is maintained. The slope in the linear fit is slope: 0.50 (+/- 0.01).

**Visible experimental spectra with fiber lasers**
Photon-counting dual-comb spectroscopy can be performed with any type of frequency comb generators. We illustrate that, even with fiber lasers attenuated to only ten million clicks per second, the spectral information in amplitude and in phase, of an absorbing sample can be precisely retrieved from the photon-counting statistics.
A pair of frequency-doubled erbium-doped femtosecond mode-locked fiber lasers generate two train of optical pulses at a central frequency of 384 THz (Supplementary Fig. 4, Methods). Their repetition frequencies are approximately 100 MHz and exhibit a difference of $\delta f_{rep}$ equal to -12.5 kHz. One fiber comb is self-referenced against a radio-frequency clock. A feed-forward dual-comb technique [35] establishes mutual coherence between the two frequency combs, effectively stabilizing their relative phase and timing fluctuations. The phase-matching properties of the periodically-poled lithium niobate crystals limit the spectral span to roughly 0.12 THz. The self-referenced comb passes through a Rb vapor cell, after which it is combined on a beam-splitter with the second comb. At one of the outputs of the beam-splitter, the beam is attenuated to a detected count rate of $8.4 \times 10^6$ clicks s$^{-1}$, which corresponds to less than one click every twenty laser pulses. The calculated power per comb before the detector is $1.5 \times 10^{-12}$ W, $10^8$ times weaker than commonly used in dual-comb spectroscopy. The average power per comb line is estimated to $1.2 \times 10^{-15}$ W. The signal is captured by a photon-counting detection module. The second output of the beam-splitter produces the trigger signal; the dual-comb interference is detected by a silicon photodiode. The fringe occurring at zero delay acts as a trigger of the multiscaler, which adds up the subsequent scans.

The multiscaler directly accumulates the time scans, each comprising $5 \times 10^5$ time bins over a duration of 3.28 ms. Over a total duration of 4,592 seconds, the interference signal gradually builds up in the counting statistics. The final interferogram time trace (Supplementary Fig. 5) consists of 41 identical bursts occurring periodically with a period $1/\delta f_{rep}$ of $8 \times 10^{-5}$ s. The Fourier transform of the interferogram reveals a transmittance spectrum with resolved comb lines (Fig. 5). The spectrum spans 0.12 THz with 1,200 comb lines well above the noise level (Fig. 5a). The Doppler-broadened $5S_{1/2}$-$5P_{3/2}$ transitions in $^{85}$Rb and $^{87}$Rb, are sampled by the comb lines of 100 MHz spacing (Fig. 5b). The comb lines exhibit a transform-limited width (Fig. 5c). The average signal-to-noise ratio of the absorption baseline is 67, whereas the calculated quantum-noise limited signal-to-noise ratio over 1,200 comb lines is 69. The phase spectrum is simultaneously retrieved. The absorption and dispersion profiles of Rb transitions, sampled by the comb lines at 100-MHz resolution, show a full width at half maximum of about 600 MHz (Supplementary Fig. 6).

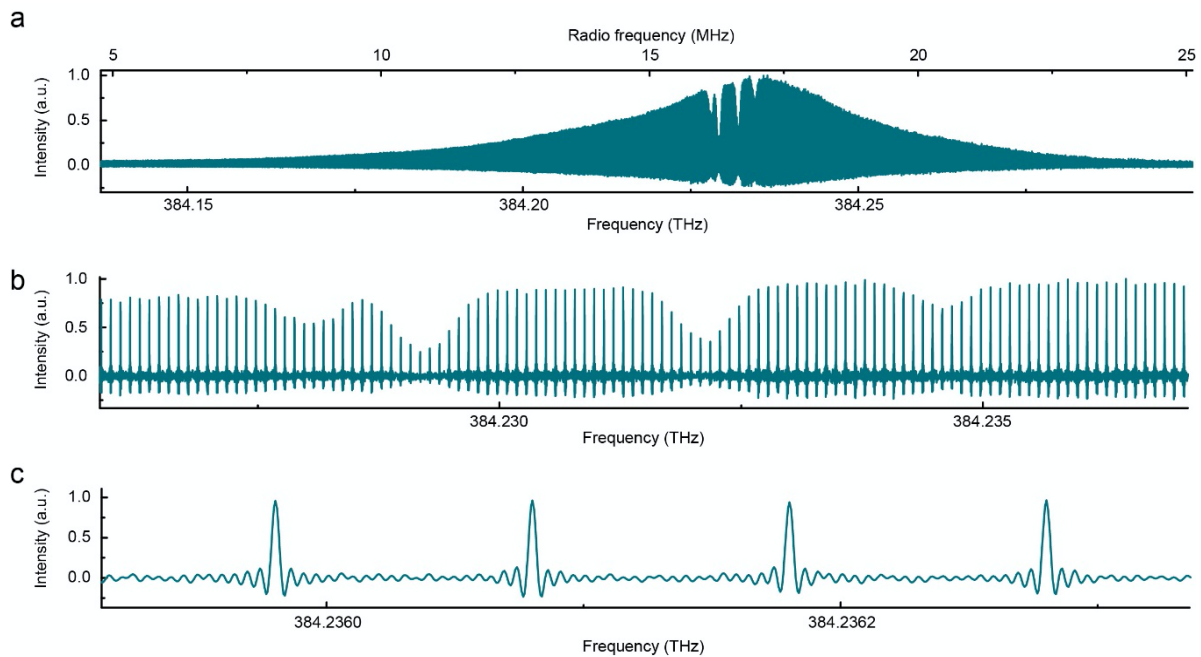

**Figure 5.   Photon-counting visible dual-comb experimental spectrum with fiber lasers at an average rate of $8.4 \times 10^6$ clicks s$^{-1}$.** The average power per comb line is $1.2 \times 10^{-15}$ W. (a) Dual-comb spectrum with 1,200 comb lines. (b) Magnified portion of (a) showing the $5S_{1/2}$-$5P_{3/2}$ transition of Rubidium sampled by the comb lines. (b) Magnified view of four comb lines. The comb lines show the expected sinc-function instrumental line shape. The full width at half-maximum of the comb lines is 365 Hz in the radio-frequency domain, corresponding to the Fourier transform limit.

## Discussion

Our experiments provide the first implementation of high-resolution linear-absorption dual-comb spectroscopy in the ultraviolet spectral range. Importantly, they also establish, with two distinct experimental setups involving frequency-comb generators of different types, that the full capabilities of dual-comb spectroscopy are extended to starved-light conditions, at power levels more than a million-fold weaker than those commonly employed in dual-comb spectroscopy. By repeatedly reaching a quantum-noise limited signal-to-noise ratio, an optimal use of the light available for the experiments is achieved. Our interferometer at the photon-level accurately reproduces the statistics of photon-counting, as shown with this signal-to-noise ratio at the fundamental limit, this instrumental line-shape that follows the theoretical expectation and this direct referencing of the frequency scale to a radio-frequency clock. The prospect of harnessing dual-comb spectroscopy at very-low light levels may be seen as counter-intuitive [33]. Here we have experimentally realized such a milestone, which will unlock novel applications. In this section we discuss the implications and prospects of our results.

Over the past years, dual-comb spectroscopy has emerged as a powerful technique for precise spectroscopy over broad spectral bandwidths. A technique of Fourier transform spectroscopy, it measures the time-domain interference between two frequency combs of slightly different repetition frequencies and reveals the spectrum through harmonic analysis of this interference pattern. However, as a technique relying on frequency measurements rather than on wavelength determination, it does not encounter the geometric limitations of state-of-the-art interferential or diffractive spectrometers. It promises a yet-unexploited potential for high precision and high accuracy. While dual-comb spectroscopy has been mostly employed for linear absorption of Doppler-broadened or collision-broadened transitions of small molecules in the gas phase, action spectroscopy has been successfully demonstrated [21,25,26] with e.g. Doppler-free two-photon excitation [25]. Nevertheless, the technique currently uses intense laser beams of an average power higher than several tens of microwatts at the detector and it interrogates samples of millions of atoms or molecules.

In a broad context though, measuring spectroscopic signatures at extremely-low light levels has become crucial in many areas of science and technology. This ranges from precision spectroscopy -where few atoms or molecules are observed in controlled conditions where the systematic effects due to interaction between atoms and between atoms and light field are minimized- to light-damageable strongly-scattering bio-medical tissues to environmental sounding of the atmosphere over long distances. Our work opens up the prospect of dual-comb spectroscopy at low light levels in these challenging scenarios, which will leverage the host of specific features of the technique.

An exciting forthcoming application is certainly the development of dual-comb spectroscopy in the short-wavelength range, which would unlock precise vacuum- and extreme-ultraviolet molecular spectroscopy over broad spectral spans. Currently broadband UV spectroscopy is limited in resolution and accuracy [7,36,37] and, at short-wavelengths, it relies on unique instrumentation at a unique facility [38,39]. Ultraviolet dual-comb spectroscopy is a highly sought-after objective, but a challenging one for three primary reasons. Firstly, similar to other lasers, frequency comb sources that emit directly in the ultraviolet region are not readily available. The current approach for generating ultraviolet frequency comb radiation involves nonlinear frequency conversion through harmonic generation in crystals (or rare gas jets for shorter wavelengths). However, this process leads to low conversion efficiencies and consequently weak powers. Secondly, during this conversion process, the phase noise of the comb source is multiplied, typically by the harmonic number. Lastly, dual-comb spectroscopy relies on interferometry. Many interferometry applications necessitate defining the fringes within a hundredth of the wavelength. Achieving such a precision in an interferometer operating at short wavelengths is technically challenging, and, for dual-comb spectroscopy, it becomes even more demanding due to the adverse effect of phase-noise multiplication resulting from nonlinear frequency conversion.

Efficient solutions have been reported in other contexts for generating low-noise combs in the ultraviolet and for building dual-comb interferometers with long coherence-times. Building on these provides a starting point. Conversely, the pressing issue of handling low powers in dual-comb spectroscopy remained unexplored until our work. In [4], fluorescence excitation of a rare gas by a single comb line of a power of 10 pW unlocked linear direct frequency-comb spectroscopy at 63 nm. Our present concept at the femtowatt level will enable to extend such a pioneering experiment to broadband dual-comb excitation spectroscopy [25] in the XUV, to measure at even lower powers thus at shorter wavelengths and potentially to acquire linear-absorption broadband spectra. Furthermore, the generation of frequency combs at frequencies higher than 1,600 THz (wavelengths shorter than 180 nm) has relied on cavity-enhanced high-harmonic generation in a rare gas, a complex process [40] whose use in dual-comb spectroscopy is daunting. Our work opens up alternative strategies: single pass high-harmonic generation in a rare gas [41] or the emerging high-harmonic generation in solids [42] become conceivable options, even at high repetition frequencies.

**Acknowledgments.** We thank Karl Linner for technical support.
**Funding.** European Union (ERC Advanced Grant, COMB, project 101054704). Max-Planck Society. Munich Center for Quantum Science and Technology funded by the Deutsche Forschungsgemeinschaft (DFG, German Research Foundation) under Germany's Excellence Strategy – EXC-2111 – 390814868. Carl-Friedrich von Siemens Foundation.

**Methods**

**Near-ultraviolet frequency comb generation, photon-counting dual-comb interferometer and experimental spectra.**

A continuous-wave (cw) extended-cavity widely-tunable diode laser emits at a center frequency of 193 THz (1550 nm) with an average power of 40 mW. The cw-laser beam is split into two beams. In one beam, the frequency of the cw-laser is shifted using an acousto-optic modulator [11,19,43,44], to avoid aliasing in the dual-comb interferograms. Here the frequency shift is $f_{AOM}$=40 MHz. Each beam is modulated by an electro-optic amplitude modulator followed by an electro-optic phase modulator. The phase modulator is driven at a voltage of about 4.4 $V_\pi$ where $V_\pi$ is the voltage required for inducing a phase change of π. The amplitude modulator gates the linear part of the up- or down-chirp induced by the phase modulator, leading to a flat-top spectral intensity distribution. For driving repetition frequencies at around 500 MHz, about 27 comb lines are generated within a 3-dB bandwidth. Each near-infrared beam is amplified up to 400 mW using an erbium-doped fiber amplifier and it is frequency-doubled to 384 THz (780 nm) in a 40-mm-long periodically-poled lithium-niobate crystal. The residual 192-THz light is filtered out with a dichroic mirror and the 384-THz beam is focused onto a 10-mm-long $BiB_3O_6$ (BIBO) crystal. Two near-ultraviolet combs, each of 100 lines and of repetition frequency $f_{rep}$ and $f_{rep}+\delta f_{rep}$ respectively, are generated at around 772 THz. In these experiments, we tune the center frequency of the combs between 770 and 774 THz but the specifications of the involved instrument allow for tunability between 750 and 784 THz. Typically, we chose $f_{rep}$ between 200 MHz and 500 MHz and $\delta f_{rep}$ between 1.6 MHz and 500 kHz. The span is 50 GHz at a line spacing of 500 MHz (Figs.2,3), and 26 GHz at a line spacing of 200 MHz (Supplementary Fig.3). The detected rate of each comb is at most 2.5 $10^7$ counts $s^{-1}$, corresponding to an average power for each comb beam up to 5 $10^{-11}$ W.

The two near-ultraviolet beams are superimposed on a beam splitter to form an interferometer. At one of the outputs of the beam splitter, an optical short-wavelength-pass filter further selects the ultraviolet light. The beam passes through a heated caesium-vapor cell and it is detected by a photon-counting detector. The detector is a photomultiplier of a single-electron response width of 600 ps and a quantum efficiency of 25% at optical frequencies around 772 THz. Photon rates of several $10^7$ photons $s^{-1}$ are detected, corresponding to optical powers before the detectors in $10^{-11}$ W. The clicks are counted as a function of time by a multiscaler. The multiscaler is started with a trigger signal generated by a frequency division of the 10-MHz clock that synchronizes all electronics in the experiment. Typically, for $f_{rep}$ = 500 MHz and $\delta f_{rep}$ = 1.6 MHz, we set the trigger to 200 kHz. At each trigger signal, the multiscaler adds up the clicks to those of previous scans over a total duration of 5 μs, with a time resolution of 160 ps. By accumulating photon-count statistics, an interferogram is reconstituted. The interferograms are zero-filled with a factor of 16 (Fig.2a,b) or 8 (Fig.2c, Fig.3a,c). A complex Fourier transform of the interferogram reveals the amplitude spectrum.

In Fig. 2a, the count rate is 2.7 $10^7$ counts $s^{-1}$. The strong line at around 770.71 THz (at a radio frequency of 80 MHz) is an artifact corresponding to the frequency-doubled frequency shift of the acousto-optic modulator.

In the experiments in Fig. 3, a caesium vapor cell of a length of 75 mm is employed. For the weak $6S_{1/2}$-$8P_{1/2}$ transition at around 770 THz, the cell is heated to about 378 K (Figs.3a,b), whereas for the $6S_{1/2}$-$8P_{3/2}$ transition at around 773 THz, it is heated to about 359 K (Figs.3c,d). If the signal-to-noise ratio in Figs.3b,d is normalized to an accumulation time of 1 second, one obtains a signal-to-noise ratio at 1 second on the order of 20 $s^{-1/2}$ at an average power per comb of 30 pW before detection.

For $f_{rep}$ = 200 MHz, we set $\delta f_{rep}$ = 0.5 MHz and the frequency of the trigger to 100 kHz. The spectrum (Supplementary Fig. 3) spans over 26 GHz and includes more than 130 comb lines.

While the spectral line shapes are more densely sampled at 200-MHz line spacing the signal-to-noise ratio is 260 at an accumulation time of 328 s.

**Visible-range photon-counting dual-comb interferometer and experimental spectra.**
Two erbium-doped-fiber mode-locked lasers of a repetition frequency of 100 MHz emit at 192 THz (Supplementary Fig.4). The repetition frequency $f_{rep}$ = 100 MHz and the carrier-envelope offset frequency $f_{ceo}$ of the first comb laser (called master comb generator) are stabilized against the radio-frequency signal of a hydrogen maser, using self-referencing with a *f-2f* interferometer. The second comb (called slave comb) has a repetition frequency $f_{rep}$+ $\delta f_{rep}$ with $\delta f_{rep}$=−12.5 kHz and a carrier-envelope offset frequency $f_{ceo}$ + $\delta f_{ceo}$. It is stabilized against the first comb through feed-forward control of the relative carrier-envelope offset frequency. This scheme allows long coherence times for the interferometer and direct averaging of the time-domain interferograms over more than one hour. The feed-forward control scheme, which uses an external acousto-optic modulator, has been described in detail in Ref. [35]. In addition, as one comb is fully referenced to a radio-frequency clock, absolute calibration of the frequency scale is directly achieved.

In our setup, each laser beam is frequency-doubled to 384-THz in a 40-mm-long periodically-poled lithium-niobate crystal. The span is limited to 100 GHz by the long crystals to reduce the volume of data, and to adapt to the capabilities of our multiscaler. At the output of the periodically-poled lithium-niobate crystals, dichroic mirrors filter out the 192-THz radiation. The 384-THz beam of the master comb generator passes through a 3-cm-long cell with rubidium in natural abundance. The cell is heated to 315.5 K. The beam is then combined on a beam-splitter with the beam of the second comb generator. One output of the beam-splitter is attenuated to an average optical power of $3 \times 10^{-12}$ W. It is detected by a fiber-coupled single-photon counting module based on an avalanche photodiode. The counting module detects an average rate of $8.4 \times 10^6$ clicks per second. The detection efficiency of the module is about 70%. The second output of the beam-splitter is detected by a fast silicon photodiode and the central interference fringe provides a trigger signal. The counts of the single-photon counting module are acquired by a multiscaler triggered by the the fast silicon photodiode. The sampling rate of the multiscaler is $156.25 \times 10^6$ samples s$^{-1}$. On average, one click is detected every twelfth laser pulses. As many as $1.4 \times 10^6$ triggered sequences, each of 3.28 ms, are summed up to provide, from the photon counting statistics, a time-domain interference signal (Supplementary Fig.5) accumulated over a total time of 4,592 s. The interferogram comprises 41 individual interferograms that recur at a period of $1/\delta f_{rep} = 8 \times 10^{-5}$ s for a total of 512,500 samples, currently limited by the multiscaler capabilities.

The raw interferometric signal shows a significant non-interferential part (Supplementary Fig.5a): owing to dark counts of the detector, stray light, parasitic light leaking through the fiber before the counting module, slightly spectral-envelope shapes of the combs, the fringe visibility *V* is 36%. The amplitude of the 41 zero-optical-delay bursts remains constant, illustrating that the sequences are efficiently averaged over the time of the experiment. The bandwidth of the electronics does not entirely filter out the pulses at the detector, explaining the residual pulse pattern which is maximum around zero optical delay and minimum at the largest optical delay, in the middle of the interferometric sequence. Simple numerical filtering returns the usual interferogram shape (Supplementary Fig.5b), where the modulation due to the absorbing rubidium is clearly visible even in the region of the largest optical retardations of 5 ns (Supplementary Fig.5c). The complex Fourier transform of the interferogram (Supplementary Fig.5b) provides the complex response (amplitude and phase) of the sample. The unapodized amplitude spectrum, interpolated through four-fold zero-filling of the interferogram, shows well-resolved comb lines with the imprint of the Doppler-broadened 5S$_{1/2}$-5P$_{3/2}$ transitions in $^{85}$Rb and $^{87}$Rb (Fig.5). The individual comb lines show the instrumental line-shape of the interferometer, a cardinal sine, which is induced by the finite measurement time. The full-width at half-maximum of the spectral envelope is 38.4 GHz, corresponding to 384 comb lines spaced by 100 MHz. Owing to the good signal-to-noise ratio, more than 1,200 are measurable in the spectrum. Sampling the spectra at the comb line positions reveals the transmittance and dispersion spectra of the Rb transitions (Supplementary Fig.6).

**Derivation of the quantum-noise-limited signal-to-noise ratio in photon-counting mode:**
We adapt the formalism developed in [45] to our experimental situation. We consider a dual-comb interferometer, where only one output of the interferometer is detected by a photon counter. The time-domain interferogram is composed of a sequence of $L$ individual interferograms. An individual interferogram spans over laboratory times between -1/(2δ$f_{rep}$) to +1/(2δ$f_{rep}$), corresponding to optical delays ranging between -1/(2$f_{rep}$) to +1/(2$f_{rep}$). As explained above, an individual interferogram is acquired over an accumulation time $T_{indiv}$, resulting of the addition, for each time bin, of photon counts over several triggered scans in order to statistically reconstitute the individual interferogram. Assuming that sufficient statistics has been accumulated, by a proper selection of the sampling and comb parameters, all the individual interferograms are expected to be identical, but for the noise.

At zero optical delay ($t = 0$), the quantum-noise-limited signal-to-noise ratio in one individual interferogram is given by

$$\left(\frac{S}{N}\right)_{t=0} = \frac{n_{interf}}{\sqrt{n+n_{interf}}}$$

where $n$ is the number of detector clicks corresponding to non-interferometrically-modulated signal, accumulated for the time bin at zero optical delay over an integration time of $1/f_{rep}$ and $n_{interf}$ is the number of clicks that contribute to the interferometric signal for the same time bin (that is, the total number of clicks minus the number of clicks for non-interferometric contributions).

In an ideal interferometer, one would expect $n_{interf} = n$, leading to $\left(\frac{S}{N}\right)_{t=0} = \sqrt{\frac{n}{2}}$. Experimentally though, many factors contribute to deviations in an additive fashion or in a multiplicative one. Additive contributions include residual stray light or the dark counts of the photon counter. Multiplicative contributions can be due to optical misalignment; the beam-splitter may not have the optimal reflection and transmission coefficient; the two interfering combs may not be identical: they may exhibit different power, spectral intensity distribution, polarization, etc.

The fringe visibility $V$ can conveniently be introduced: $V = \frac{n_{interf,max} - n_{interf,min}}{n_{interf,max} + n_{interf,min}}$, with $n_{interf,max}$ and $n_{interf,min}$ being the maxima and minima of the interference counts, respectively. In an ideal interferometer, $V = 1$. In our experiments, the additive noise is negligeable.

In such a case, $n_{interf,max} = n + n_{interf}$ and $n_{interf,min} = n - n_{interf}$, thus $n_{interf} = V n$. Consequently, the signal-to-noise ratio at zero optical delay can be written:

$$\left(\frac{S}{N}\right)_{t=0} = \frac{V}{\sqrt{1+V}}\sqrt{n},$$

The signal-to-noise ratio $\left(\frac{S}{N}\right)_\nu$ t the frequency ν in the spectrum is related to the signal-to-noise ratio $\left(\frac{S}{N}\right)_{t=0}$ at zero optical delay in the time-domain interferogram by the expression:

$$\left(\frac{S}{N}\right)_\nu = \sqrt{\frac{2}{K} \frac{B(\nu)}{\overline{B_e}}} \left(\frac{S}{N}\right)_{t=0}$$

where $B(\nu)$ is the spectral distribution at the frequency $\nu$. $\overline{B_e}$ is the mean value of the spectral function $(B_e(\nu) = \frac{1}{2}(B(\nu) + B(-\nu))$, which accounts for the unphysical negative frequencies. The number of time bins $K$ in the interferogram is twice the number of spectral elements in the actual spectral distribution which – according to our sampling conditions here - spans from frequency 0 to frequency $f_{rep}/2$ on the radio-frequency scale.

Under the simplifying hypothesis that the spectrum is made of $M$ comb lines, all of equal intensity, one can express the ratio $B(\nu)$ to $\overline{B_e}$ at a frequency $\nu$ corresponding to a comb line position:

$$\frac{B(\nu)}{\overline{B_e}} = \frac{1/M}{1/K} = \frac{K}{M}$$

Consequently,

$$\left(\frac{S}{N}\right)_\nu = \frac{\sqrt{2K}}{M}\left(\frac{S}{N}\right)_{t=0} = \sqrt{2}\frac{V}{\sqrt{1+V}}\frac{\sqrt{K}}{M}\sqrt{n}$$

Upon summing up $L$ individual interferograms, the quantum-limited signal-to-noise ratio at the comb line positions becomes:

$$\left(\frac{S}{N}\right)_\nu = \sqrt{2}\frac{V}{\sqrt{1+V}}\frac{\sqrt{K}}{M}\sqrt{n\,L} \qquad (1)$$

Eq. (1) can also be written using the detected photon rate $N_{phot}$ (in photons s$^{-1}$) in the interferogram:

$$n = \frac{N_{phot}\, T_{indiv}}{K}$$

where $K$ is the number of time bins in the individual interferogram and $T_{indiv}$ is the accumulation time for one individual interferogram which optical delays $-1/(2f_{rep})$ to $+1/(2f_{rep})$ with time bins of $1/f_{rep}$.

Eq. (1) becomes

$$\left(\frac{S}{N}\right)_\nu = \sqrt{2}\frac{V}{\sqrt{1+V}}\frac{1}{M}\sqrt{N_{phot}\, T_{indiv}\, L} \qquad (2)$$

Moreover, the measurement of $N_{phot}$ in the interferogram enables to infer the average power $P$ before the photon counter.

$$N_{phot} = \frac{P\, QE}{h\nu}$$

where $P$ is the average optical power before the counter and QE is the counter quantum efficiency.

One can also write the quantum-limited signal-to-noise ratio at the optical frequency $\nu$ of a comb line as an equation involving the average power rather than the photon counts:

$$\left(\frac{S}{N}\right)_\nu = \frac{\sqrt{2}}{M}\frac{V}{\sqrt{1+V}}\sqrt{\frac{P\, QE}{h\nu}\, T_{indiv}\, L} = \left(\frac{S}{N}\right)_\nu = \frac{\sqrt{2}}{M}\frac{V}{\sqrt{1+V}}\sqrt{\frac{P\, QE}{h\nu}\, T_{tot}}$$

where $T_{tot} = T_{indiv}\, L$ is the total accumulation time in the entire recording.

**Comparison measured and calculated signal-to-noise ratio in photon-counting mode:**

|  | $M$ | $V$ | $T_{indiv}$ (s) | $L$ | $T_{tot}$ (s) | $N_{phot}$ (s$^{-1}$) | Calculated power per comb before detection (W) | Experimental SNR | Calculated SNR | Trigger frequency (kHz) |
|---|---|---|---|---|---|---|---|---|---|---|
| Fig.3b Fig.4b | 100 | 0.33 | 33.4 | 8 | 270.4 | 2.98 10$^7$ | 30 10$^{-12}$ | 345 | 360 | 200 |
| Fig.3d | 100 | 0.41 | 20.6 | 4 | 82.4 | 2.83 10$^7$ | 30 10$^{-12}$ | 205 | 235 | 400 |
| Fig. 5 | 1,200 | 0.36 | 112 | 41 | 4,592 | 8.4 10$^6$ | 1.5 10$^{-12}$ | 67 | 69 | 0.3 |

**Table 1. Experimental parameters and calculated quantum-noise limited signal to noise ratio.** In the experiments of Figs. 3b, 4b, 3d, $f_{\text{rep}}$ =500 MHz, δ$f_{\text{rep}}$ =1.6 MHz. The sampling rate $f_{\text{rep}}$ and the integration time per time bin is 1/$f_{\text{rep}}$. The calculated signal-to-noise ratio (SNR) is calculated using Equation 2. We attribute the slight mismatch between the measured and calculated by the rudimentary model that we use, which neglects additive noise sources and assumes that all comb lines have the same intensity. The calculated average power per comb is half the total power before detection and includes the photon-counter efficiency (25% in Fig. 3,4, 70% in Fig. 5).

**Supplementary Figures**

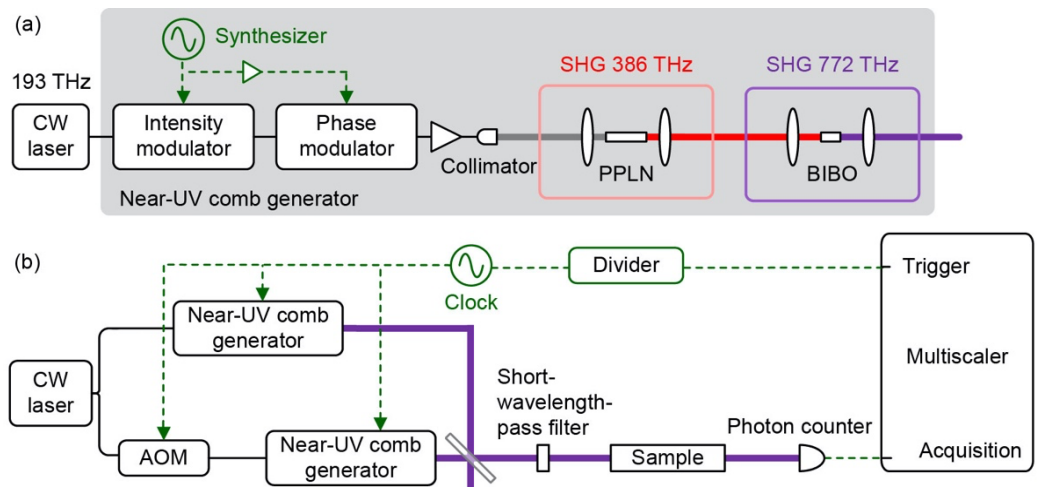

**Supplementary Figure 1: Detailed experimental setup for near-ultraviolet photon-counting dual-comb spectroscopy.** (a) A near-infrared continuous-wave laser is modulated by electro-optic modulators and frequency-doubled twice to generate a weak comb at 772 THz with an average power of a few $10^{-11}$ W. (b) Two combs are generated as sketched in (a) from a single continuous-wave laser and are combined on a beam-splitter. One output of the beamsplitter is counted by a photon counter and accumulated as a function of arrival time after the trigger by a multi-scaler. Typically, a few $10^7$ clicks per second are recorded, which corresponds to less than a click per comb repetition-frequency occurrence.

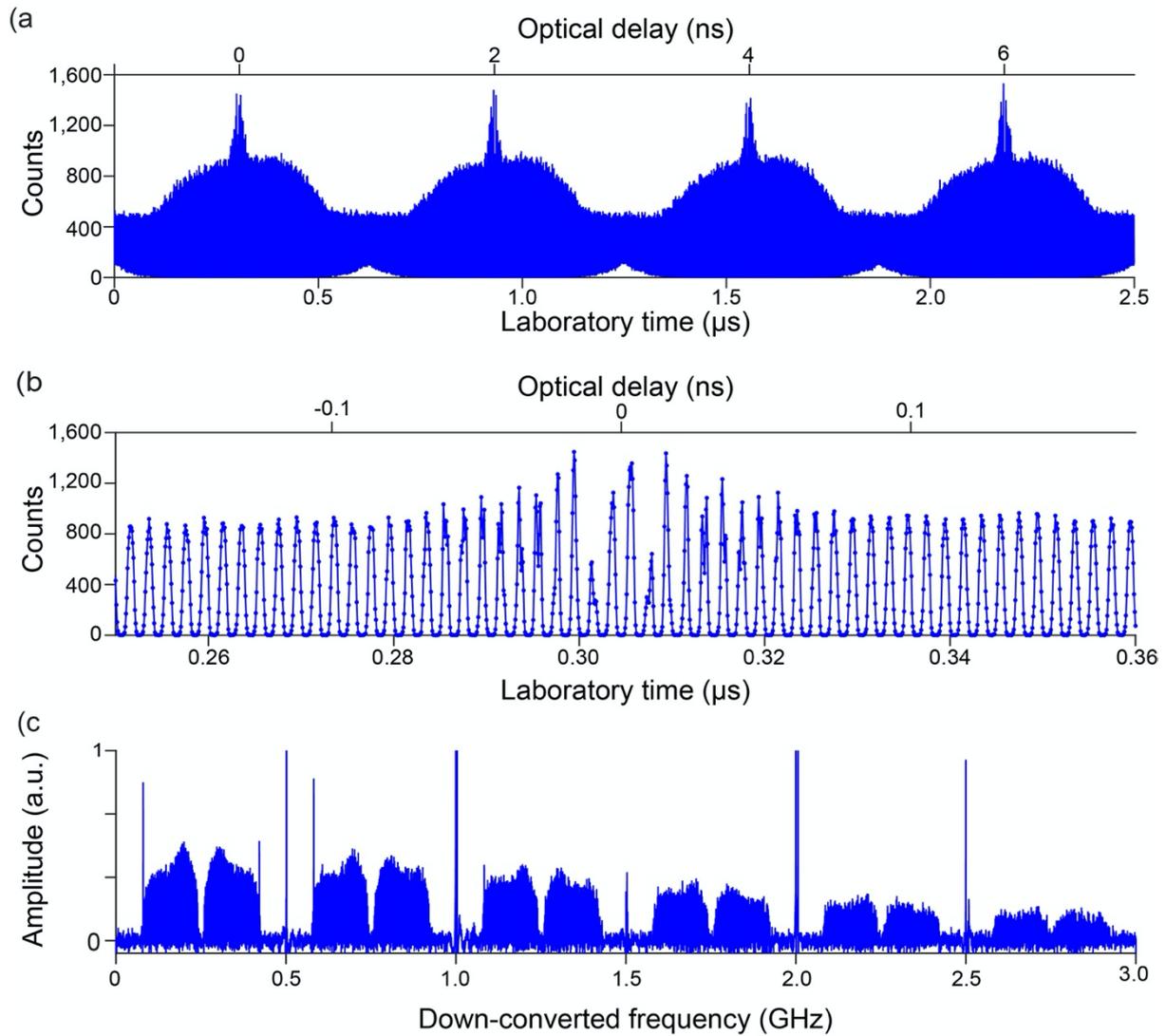

**Supplementary Figure 2: Photon-counting near-ultraviolet interferogram and spectrum at a sampling rate of 12.5 $10^9$ bins per second.** The count rate is 5 $10^7$ counts s$^{-1}$ and the accumulation time is 0.2 s. (a) The interferogram features $\delta f_{rep}$=1.6 MHz and a trigger rate of 200 kHz. (b) Magnified portion of (a) in the region of zero optical delay. (c) The Fourier transform of the interferogram reveals the multiple aliases of the spectrum (12 in the representation) with decreasing intensity as the detection bandwidth drops.

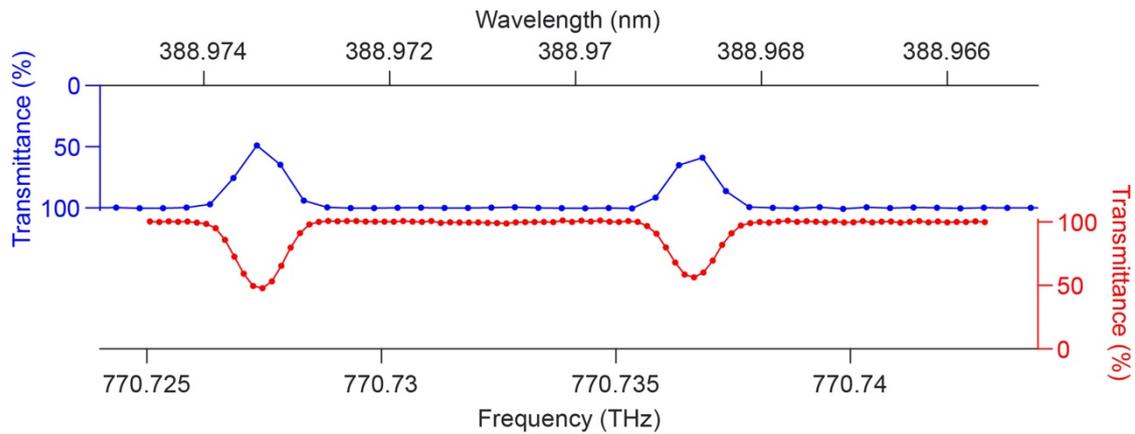

**Supplementary Figure 3. Repetition-frequency agility in photon-counting dual-comb spectroscopy.** Experimental transmission spectra of the caesium $6S_{1/2} \rightarrow 8P_{1/2}$ transitions at a temperature of 378 K with a resolution of 500 MHz (blue, also in Fig.3a,b) and of 200 MHz (red). The resolution is freely chosen by adjusting the comb line spacing and resolving the comb lines in the spectra. The spectrum at 200-MHz comb line spacing results from a count rate of $2.3 \times 10^7$ counts s$^{-1}$, corresponding to an average power of 23 pW per comb before the detector. The accumulation time is 328 s.

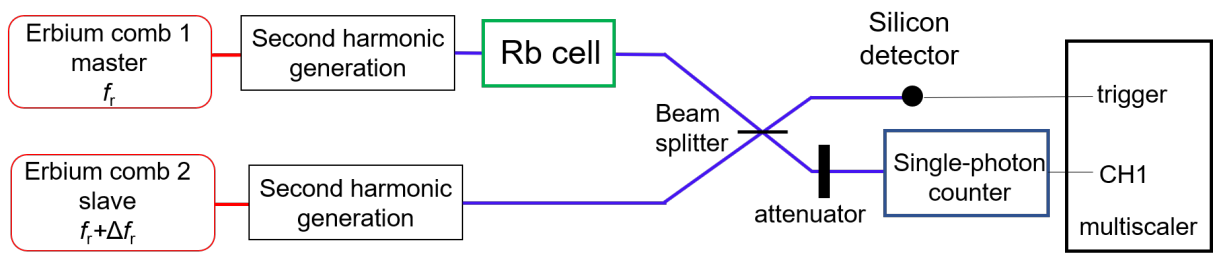

**Supplementary Figure 4. Experimental setup of visible photon-level dual-comb spectroscopy.** Two frequency combs based on frequency doubled fiber lasers form a dual-comb interferometer at 384 THz. The beam of one comb interrogates a rubidium vapor cell. One output of the interferometer is strongly attenuated to an average power of $3 \times 10^{-12}$ W, corresponding to one click every twenty-four laser pulses on average. The photon clicks are counted by a photon-counter and a multiscaler accumulates them as a function of optical delay.

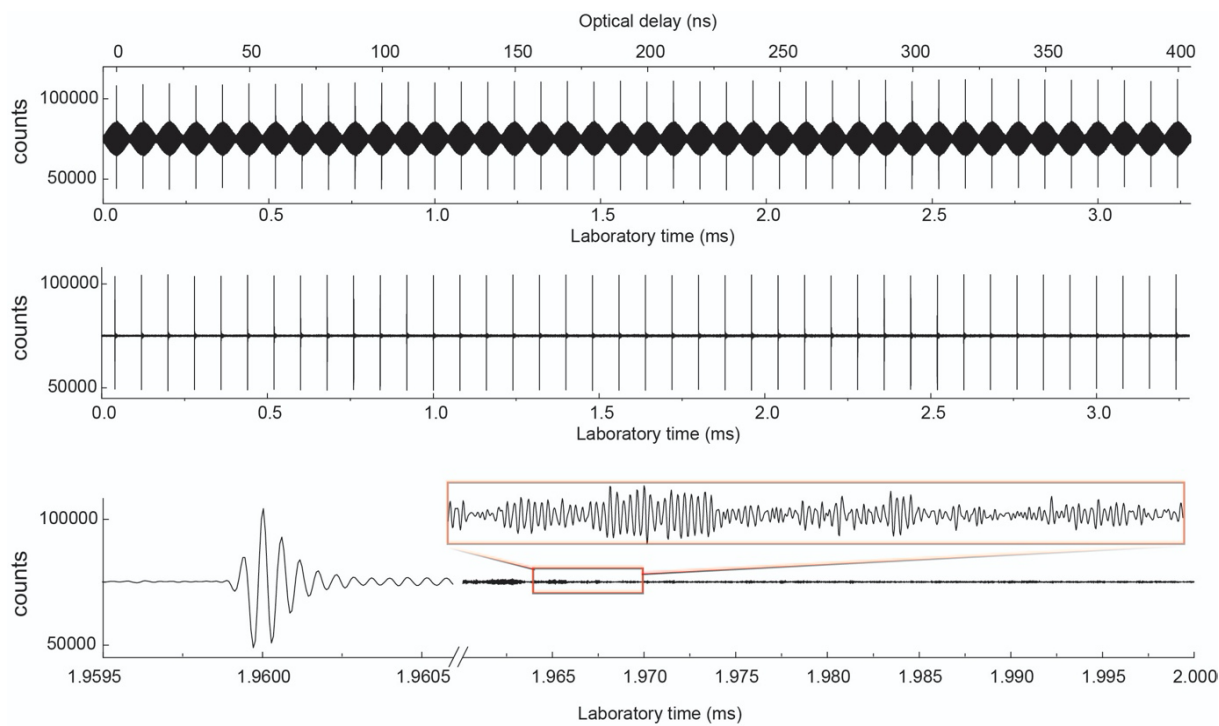

**Supplementary Figure 5. Photon-counting dual-comb interferogram with fiber-laser systems at 384 THz.** The time-domain trace results from $1.4 \times 10^6$ scans, each with a duration of 3.28 ms. The total measurement time is 76.5 min at a counting rate of $8.4 \times 10^6$ clicks s$^{-1}$. (a) Experimentally acquired interferogram of the counting statistics, where the 100-MHz laser pulse trains are partly resolved. (b) The interferogram of (a) is numerical filtered out with a 40-MHz low-pass filter. (c) Magnified view of a portion of the dual-comb interferogram in (b) showing the interferometric modulation due to the rubidium transitions.

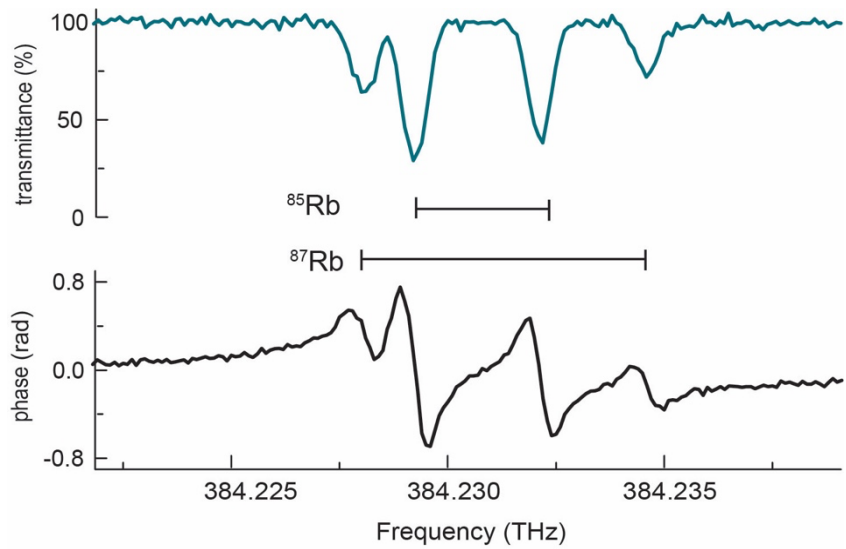

**Supplementary Figure 6. Photon-counting dual-comb transmittance and phase spectra in the region the 5S$_{1/2}$-5P$_{3/2}$ of Rubidium.** The spectrum is sampled by the individual comb lines at a resolution of 100 MHz. The spectrum is measured at an average power of 3 10$^{-12}$ W and the signal-to-noise ratio is limited by the quantum noise. For each isotope ($^{85}$Rb or $^{87}$Rb), the resolved hyperfine splitting of the ground state 5S$_{1/2}$ leads to two transitions per isotope.